\documentclass[runningheads]{llncs}
\usepackage{algorithm2e}

\usepackage{xcolor}

\usepackage{amsmath, amssymb}
\usepackage{amsthm}

\usepackage[T1]{fontenc}
\usepackage{graphicx}
\begin{document}
\title{VLWE: Variety-based Learning with Errors for Vector Encryption through Algebraic Geometry}

\author{Dongfang Zhao} 
\institute{
University of Washington, USA \\
\email{dzhao@cs.washington.edu}
}

\maketitle              % typeset the header of the contribution
\begin{abstract}
Lattice-based cryptography is a promising foundation for post-quantum secure cryptographic schemes, with the Learning with Errors (LWE) problem serving as a cornerstone for key exchange, encryption, and homomorphic computation. Existing structured variants of LWE, such as Ring-LWE (RLWE) and Module-LWE (MLWE), rely on polynomial rings to improve efficiency. However, these constructions inherently follow traditional polynomial multiplication rules and are limited in their ability to represent structured vectorized data. This work introduces Variety-LWE (VLWE), a new class of structured lattice problems built upon algebraic geometry. Unlike RLWE and MLWE, which use polynomial quotient rings with standard multiplication, VLWE operates over multivariate polynomial rings defined by algebraic varieties. A key distinction is that these polynomials do not contain mixed variables, and the multiplication operation is defined coordinate-wise rather than via standard polynomial multiplication. This structure enables direct encoding and homomorphic processing of high-dimensional data while maintaining worst-case to average-case hardness reductions. We establish the security of VLWE by reducing it to solving multiple independent instances of Ideal-SVP, demonstrating its resilience against classical and quantum attacks. Furthermore, we analyze the impact of hybrid algebraic-lattice attacks, showing that existing Gröbner basis and lattice reduction techniques do not directly compromise VLWE’s security. Building on this foundation, we construct a vector homomorphic encryption scheme based on VLWE, which supports structured computations while maintaining controlled noise growth. This scheme offers potential advantages for privacy-preserving machine learning, encrypted search, and secure computations over structured data. Our results position VLWE as a novel and independent paradigm in lattice-based cryptography, leveraging algebraic geometry to enable new cryptographic capabilities beyond traditional polynomial quotient rings.
\keywords{Lattice-Based Cryptography \and Homomorphic Encryption \and Algebraic Geometry.}
\end{abstract}

\section{Introduction}

Lattice-based cryptography has become one of the most fundamental components of post-quantum cryptographic security due to its worst-case to average-case hardness guarantees and its applicability to a wide range of cryptographic constructions. The Learning with Errors (LWE) problem serves as a cornerstone of lattice-based cryptography, providing the foundation for fully homomorphic encryption (FHE), digital signatures, and key exchange protocols. Structured variants of LWE, such as Ring-LWE (RLWE) and Module-LWE (MLWE), improve efficiency by leveraging polynomial rings, but they inherently follow traditional polynomial multiplication rules and are limited in their ability to capture structured high-dimensional data representations.

Existing techniques for handling vectorized computations in lattice-based cryptography typically rely on two approaches: batching methods and repeated independent encryptions. Batching, as employed in certain FHE schemes, allows encoding multiple scalar values into a single ciphertext by utilizing the Chinese Remainder Theorem (CRT) over cyclotomic rings. However, this approach is inherently constrained by the number of available CRT slots, which depends on the ring dimension and modulus, limiting its flexibility for arbitrary high-dimensional data processing. Alternatively, performing independent encryptions for each component of a vector introduces significant ciphertext expansion and computational overhead, particularly in homomorphic operations where error growth accumulates across multiple ciphertexts.

In this work, we introduce Variety-LWE (VLWE), a new structured lattice problem constructed from algebraic geometry. Unlike RLWE and MLWE, which rely on polynomial quotient rings with standard multiplication, VLWE operates over multivariate polynomial rings defined by algebraic varieties. A key distinction is that the defining polynomials in VLWE do not contain mixed variables, and the multiplication operation is defined coordinate-wise rather than through standard polynomial multiplication. This fundamental difference alters the algebraic structure of the underlying lattice problem, leading to new computational properties and hardness assumptions.

We formally define VLWE and establish its cryptographic hardness by demonstrating worst-case to average-case reductions. Furthermore, we construct a homomorphic encryption scheme based on VLWE, which enables structured computations over encrypted vector spaces while maintaining controlled noise growth. Our security analysis shows that VLWE reduces to solving multiple independent instances of Ideal-SVP, making it resistant to both classical and quantum attacks. Additionally, we examine hybrid algebraic-lattice attacks, demonstrating that Gröbner basis and lattice reduction techniques do not directly compromise VLWE’s security.

Our work positions VLWE as a novel approach to structured lattice problems, leveraging algebraic geometry to provide new cryptographic capabilities beyond traditional polynomial quotient rings. By introducing a new algebraic framework for lattice-based encryption, VLWE opens new directions for efficient and secure cryptographic protocols in high-dimensional data environments.

\subsection{Main Results}

% In this work, we introduce the Variety-LWE (VLWE) problem as a structured generalization of Ring-LWE, extending its algebraic foundation to multivariate polynomial rings defined over algebraic varieties. Our main results establish the theoretical security foundations of VLWE, analyze its computational efficiency, and demonstrate its applicability in vector homomorphic encryption.

\paragraph{Theoretical Security and Hardness Reductions~(\S\ref{sec:hardness})}
We prove the worst-case to average-case hardness of VLWE by establishing its connection to fundamental lattice problems:
\begin{itemize}
    \item \textbf{Reduction to Ideal-SVP:} We show that solving VLWE in the average case is at least as hard as solving multiple independent instances of Ideal-SVP in structured lattices.
    \item \textbf{Security Against Classical and Quantum Attacks:} VLWE inherits the worst-case hardness guarantees of RLWE while introducing additional algebraic constraints that influence cryptanalytic attack strategies.
    \item \textbf{Analysis of Hybrid Algebraic-Lattice Attacks:} We investigate the impact of algebraic techniques combined with lattice reduction methods, demonstrating that existing hybrid attacks are not directly applicable to the VLWE structure.
\end{itemize}

\paragraph{Computational Efficiency and Noise Growth~(\S\ref{sec:vlwe_err})}
The computational efficiency of VLWE is analyzed through its impact on encryption, decryption, and homomorphic operations:
\begin{itemize}
    \item \textbf{Coordinate-Wise Homomorphic Operations:} VLWE enables direct vector encryption, avoiding the need for batch encoding techniques used in RLWE-based FHE schemes.
    \item \textbf{Noise Growth Analysis:} We provide theoretical bounds on error propagation under homomorphic addition and multiplication, demonstrating that VLWE supports controlled noise growth through modulus switching and relinearization.
    \item \textbf{Asymptotic Complexity:} The computational cost of solving VLWE is shown to be at least as hard as solving structured Ideal-SVP, with complexity estimates scaling as \( \tilde{O}(n d \cdot 2^{\Omega(d)}) \), where \( d \) represents the degree of the defining variety constraints.
\end{itemize}

\paragraph{Applications in Cryptographic Protocols~(\S\ref{sec:scheme})}
We demonstrate the applicability of VLWE in various cryptographic settings:
\begin{itemize}
    \item \textbf{Vector Homomorphic Encryption:} VLWE enables efficient encryption and homomorphic evaluation of vectorized data, facilitating secure computations over high-dimensional encrypted datasets.
    \item \textbf{Post-Quantum Secure Computation:} By inheriting the security assumptions of RLWE, VLWE serves as a promising foundation for designing lattice-based protocols that remain secure against quantum adversaries.
    \item \textbf{Privacy-Preserving Machine Learning:} The multivariate polynomial structure of VLWE aligns naturally with machine learning models that operate on structured embeddings, allowing secure inference and encrypted model evaluations.
\end{itemize}

\subsection{Related Work}

\paragraph{Existing Lattice-Based Hardness Assumptions}  
Lattice-based cryptography has gained widespread attention as a post-quantum alternative due to its worst-case hardness guarantees and efficiency in cryptographic applications. The Learning with Errors (LWE) problem~\cite{oregev_jacm09} was first introduced as a computationally hard assumption with worst-case reductions to GapSVP, forming the basis for lattice-based cryptographic schemes. However, LWE suffers from large ciphertext sizes and computational inefficiencies. To address these limitations, Ring-LWE (RLWE)~\cite{LyubashevskyPR13} was introduced, leveraging polynomial rings to improve efficiency while preserving security reductions to worst-case ideal lattice problems. Subsequent extensions such as Module-LWE (MLWE)~\cite{Albrecht15} and NTRU-based cryptosystems~\cite{Hoffstein1998} have further explored structured variants of LWE to balance security and performance. Recent standardization efforts, such as Kyber~\cite{Bos2018kyber} and Dilithium~\cite{Ducas2018dilithium},have demonstrated the practicality of RLWE-based schemes for key exchange and digital signatures. More recent studies~\cite{May2023,Dottling2023,Pouly2024,Jain2024,Albrecht2023,Bai2023,Chen2023,Ducas2023,Eldar2023} have investigated the concrete security of LWE and its variants, refining worst-case reductions and assessing attack resistance under various parameter settings. A recent work by Peikert and Pepin~\cite{peikert2024} introduces a unified framework that systematically connects various algebraically structured LWE variants, including Ring-LWE, Module-LWE, Polynomial-LWE, Order-LWE, and Middle-Product LWE. The proposed Variety-LWE (V-LWE) departs from the above unified paradigm by introducing a security assumption that does not solely depend on lattice structure: we will show that attacks relying on ideal-lattice reductions, such as hybrid Gröbner basis and lattice-reduction methods, do not trivially extend to V-LWE.

\paragraph{Multivariate Polynomial Rings in Cryptography}  
The use of multivariate polynomial rings in cryptography has primarily been explored in the context of multivariate public key cryptosystems (MPKCs)~\cite{Patarin1996}. However, these approaches do not incorporate LWE-style noise distributions, which are crucial for worst-case to average-case reductions in lattice-based cryptography. Module-LWE~\cite{Albrecht15} extends RLWE by introducing module structures, but it remains within the single-variable polynomial framework and does not fully exploit the algebraic richness of multivariate polynomial rings. Newer works~\cite{jdey_acs23} examine alternative structured lattice assumptions, providing additional perspectives on the interplay between multivariate polynomials and lattice-based security. Our work introduces \textit{algebraic varieties} as a component of an LWE-based cryptographic assumption, extending structured lattice problems into a multivariate setting.

\paragraph{Security Reductions and Structured Lattice Problems}  
The security of structured lattice problems relies on worst-case hardness assumptions. The Ideal Shortest Vector Problem (Ideal-SVP)~\cite{Peikert16} has been extensively studied in the context of RLWE and Module-LWE, providing a foundation for their security reductions. Hybrid algebraic-lattice attacks, combining Gröbner basis techniques with lattice reduction algorithms, have been explored in cryptanalysis~\cite{hzhu_iwsec23}, particularly in the context of structured ring-based schemes. While Gröbner basis methods have shown effectiveness in attacking certain algebraic cryptosystems~\cite{Faugere2002}, their direct application to lattice-based cryptosystems remains constrained due to high computational costs~\cite{pravi_tecs24}. Our work explores the impact of hybrid algebraic-lattice attacks on Variety-LWE, demonstrating that existing attack methodologies do not directly compromise its security.

\paragraph{Homomorphic Encryption and Vector Encryption}  
Fully homomorphic encryption (FHE) schemes, such as BFV~\cite{bfv}, BGV~\cite{Brakerski2014}, and CKKS~\cite{ckks}, enable secure computation on encrypted data but primarily operate within single-variable RLWE-based constructions. Batching methods using the Chinese Remainder Theorem (CRT) allow vectorized computations~\cite{GentryHaleviSmart2012,Chillotti2020tfhe}, but they are constrained by modulus structure and require additional encoding layers. Our work introduces a direct multivariate polynomial structure that inherently supports vector encryption, providing an alternative approach to efficiently processing encrypted high-dimensional data. This approach aligns with research in privacy-preserving machine learning~\cite{Dowlin2017cryptonets}, encrypted search~\cite{danb_crypto04}, and secure federated learning~\cite{Phong2017}.

\paragraph{Attacks at Multivariate Polynomial LWE}
Bootland et al.~\cite{cbootland_ants20} proposed an attack on the multivariate Ring Learning with Errors (m-RLWE) problem in their 2020 paper \cite{cbootland_ants20}. This attack exploits algebraic relationships between the roots of the defining polynomials in the multivariate ring to reduce the m-RLWE problem to multiple lower-dimensional RLWE problems. However, this attack is not applicable to the proposed Variety-LWE due to key distinctions between the two problems.
Firstly, Variety-LWE is defined over a multivariate polynomial ring constructed from an algebraic variety, which possesses a more complex structure than the multivariate rings considered in \cite{cbootland_ants20}. Secondly, Variety-LWE enforces a ``no-mixed-terms'' constraint on the defining polynomials of the algebraic variety, ensuring that each polynomial depends on only one variable. This constraint guarantees coordinate-wise independence in Variety-LWE, preventing attackers from exploiting relationships between variables.

\section{Preliminaries}

\subsection{Algebraic Varieties}

An algebraic variety is a fundamental object in algebraic geometry, defined as the solution set of a system of polynomial equations, i.e., \textit{zero points}, or \textit{locus}. Given a field \( \mathbb{F}_q \) and an ideal \( I \subset \mathbb{F}_q[x_1, \dots, x_n] \), the affine variety associated with \( I \) is:
\[
V(I) = \{ (x_1, \dots, x_n) \in \overline{\mathbb{F}}_q^n \mid f(x_1, \dots, x_n) = 0, \forall f \in I \},
\]
where \( \overline{\mathbb{F}}_q \) denotes the algebraic closure of \( \mathbb{F}_q \), which consists of all elements that satisfy algebraic equations over \( \mathbb{F}_q \). The variety \( V(I) \) represents the common zero set of all polynomials in \( I \), forming a geometric structure that encodes algebraic relationships.

\paragraph{Basic Properties of Algebraic Varieties}  
Algebraic varieties exhibit several important properties that influence their structure and computational aspects:
\begin{itemize}
    \item \textit{Irreducibility}: A variety is irreducible if it cannot be expressed as the union of two strictly smaller varieties. This property corresponds to the primality of its defining ideal.
    \item \textit{Dimension}: The dimension of a variety is the maximum number of algebraically independent parameters needed to describe a generic point on it.
    \item \textit{Coordinate Rings}: The coordinate ring of \( V(I) \) is the quotient ring:
    \[
    R[V] = \mathbb{F}_q[x_1, \dots, x_n] / I.
    \]
    Elements of \( R[V] \) represent polynomial functions on \( V \), where two polynomials are equivalent if they produce the same values on all points in \( V \).
\end{itemize}
These properties provide structural insights into the behavior of polynomial equations in multivariate settings.

\paragraph{Hilbert’s Nullstellensatz and Radical Ideals}  
A fundamental result connecting algebraic geometry with commutative algebra is Hilbert’s Nullstellensatz, which formalizes the relationship between varieties and ideals. It states that if \( I \) is an ideal, then the set of polynomials vanishing on \( V(I) \) forms the radical ideal:
\[
\sqrt{I} = \{ f \mid f^m \in I \text{ for some } m > 0 \}.
\]
This correspondence plays a key role in computational applications, allowing algebraic manipulations on varieties to be translated into operations on their defining ideals.

\paragraph{Applications to Cryptography}  
Algebraic varieties provide a structured way to define mathematical objects that remain closed under polynomial operations. In the context of Variety-LWE, these structures enable computations within constrained polynomial spaces while maintaining efficient algebraic operations. The defining polynomials impose structured relationships on the computational domain, leading to controlled error growth and potential security advantages.

While more advanced tools from algebraic geometry, such as cohomology theories and sheaf structures, provide a broader perspective on algebraic structures, they are not required for the homomorphic encryption scheme developed in this work. The variety structures used here are sufficient for modeling encrypted vector operations in a multivariate polynomial setting.

\subsection{Error Distribution and Security Implications}

The security of Variety-LWE and its cryptographic applications relies on the careful selection of error distributions. The error term plays a crucial role in ensuring the hardness of the problem while maintaining decryption correctness.

\paragraph{Definition of the Error Distribution}  
In the Variety-LWE framework, the error \( e \) is sampled from a discrete Gaussian distribution \( \chi \) over the quotient ring \( R_q \), defined as:
\[
\Pr[e = x] \propto \exp\left(-\frac{\|x\|^2}{2\sigma^2}\right),
\]
where \( \sigma \) is the standard deviation of the Gaussian distribution, controlling the spread of the error.

The error is typically chosen such that:
\begin{itemize}
    \item It is large enough to ensure computational indistinguishability of Variety-LWE samples.
    \item It remains small enough to allow correct decryption after a bounded number of homomorphic operations.
\end{itemize}

\paragraph{Noise Selection Strategies for Security and Efficiency}  
The choice of noise parameters directly influences both the security and practicality of the encryption scheme. The following factors must be considered:

\begin{itemize}
    \item \textit{Security Against Lattice Attacks:} The noise level should be sufficiently large to prevent an adversary from distinguishing Variety-LWE samples from uniform noise. Low noise levels make lattice reduction attacks, such as BKZ and dual attacks, more feasible.
    \item \textit{Trade-off Between Security and Decryption Accuracy:} Choosing excessively large noise can lead to decryption errors in practical implementations, limiting the depth of homomorphic operations.
    \item \textit{Adaptation for Homomorphic Computation:} Different homomorphic encryption schemes require different noise growth tolerances. Parameters should be optimized to balance performance and correctness.
\end{itemize}

Typical choices of \( \sigma \) are based on lattice reduction hardness estimates. A common heuristic is to set \( \sigma \) such that the ratio \( q/\sigma \) is sufficiently large to prevent attacks, yet small enough to support deep homomorphic computations.

\paragraph{Ensuring Security Against Attacks}  
The error distribution plays a fundamental role in ensuring the security of Variety-LWE. Several attacks attempt to exploit insufficient noise or structured error distributions:
\begin{itemize}
    \item \textit{Distinguishing Attacks:} If the error is too small, an adversary can distinguish between encrypted and random elements in \( R_q \).
    \item \textit{Lattice Reduction Attacks:} If the noise level is low, basis reduction techniques such as BKZ can efficiently recover the secret key.
    \item \textit{Algebraic Attacks:} Given the variety structure, an attacker may attempt to exploit polynomial relations between ciphertexts if error terms are not sufficiently randomized.
\end{itemize}

\paragraph{Parameter Selection for Security}  
To ensure security, the choice of parameters \( (q, n, d, \sigma) \) must balance efficiency and resilience to known attacks:
\begin{itemize}
    \item The modulus \( q \) should be large enough to prevent decryption failures while limiting computational overhead.
    \item The error standard deviation \( \sigma \) should be chosen to maximize security against distinguishing attacks while allowing for correct decryption.
    \item The polynomial degree \( d \) must be sufficient to ensure hardness against lattice reduction attacks.
\end{itemize}

\subsection{Computational Complexity and Hybrid Algebraic-Lattice Attacks}

\paragraph{Complexity of Sampling and Encryption}  
The computational cost of generating a structured lattice-based encryption instance primarily consists of:
\begin{itemize}
    \item Generating random elements in \( R_q \), which involves polynomial arithmetic over a structured quotient ring.
    \item Computing inner products in the ring \( R_q \), requiring polynomial multiplication and modular reduction.
    \item Sampling from the discrete Gaussian distribution over \( R_q \), which dominates the computational cost in high-precision implementations.
\end{itemize}
The overall complexity of generating a single encryption sample depends on the algebraic properties of the underlying ring and polynomial structure.

\paragraph{Reduction to Lattice Problems}  
The security of structured lattice-based schemes is supported by worst-case hardness reductions. Specifically, solving these problems is at least as hard as solving multiple independent instances of Ideal-SVP. The reduction proceeds as follows:
\begin{itemize}
    \item Given a structured lattice instance, we construct a basis for the corresponding ideal lattice.
    \item The adversary attempting to solve the problem must distinguish noisy samples from uniform elements in \( R_q \).
    \item Using an oracle that distinguishes structured lattice samples, one can iteratively refine estimates of the shortest vector in an ideal lattice.
\end{itemize}
This reduction ensures that an efficient algorithm for solving these problems would imply an efficient algorithm for solving Ideal-SVP, which is conjectured to be hard in the worst case.

\paragraph{Known Cryptanalytic Attacks}  
Several attacks have been developed against lattice-based cryptographic schemes. We analyze how these methods apply to structured lattices.

\textit{Lattice Basis Reduction Attacks:}  
These methods, such as the BKZ algorithm, attempt to solve the underlying Bounded Distance Decoding (BDD) problem by reducing the basis of the corresponding lattice. The best known variants of BKZ, incorporating progressive sieving and enumeration techniques, achieve a complexity of:
\[
T_{\text{BKZ}}(n) = 2^{O(n/\log n)}
\]
in the best case, making direct basis reduction infeasible for sufficiently large \( n \).

\textit{Dual Attack:}  
The dual attack exploits the dual lattice to construct a distinguishing oracle for structured lattice problems. Given a sample set \( (A, b) \), an adversary seeks a short vector \( \mathbf{w} \) in the dual lattice such that:
\[
\langle \mathbf{w}, A \rangle \approx 0 \mod q.
\]
If such a vector can be found, it allows the adversary to filter out noise contributions, distinguishing valid samples from uniform random elements.

For structured lattice problems, the key challenge in applying dual attacks lies in the additional algebraic structure induced by polynomial constraints. While these constraints provide some structural information, they also increase the effective dimensionality of the problem, making direct dual attacks less efficient. The complexity of dual attacks in this setting is estimated to be:
\[
T_{\text{dual}}(n, d) = \tilde{O}(2^{\Omega(d)}),
\]
where \( d \) represents the degree of the defining polynomial constraints.

\paragraph{Hybrid Algebraic-Lattice Attacks}  
A more sophisticated cryptanalytic approach involves hybrid methods that combine algebraic techniques (e.g., Gröbner basis computations) with classical lattice reduction methods. These attacks exploit structured polynomial equations inherent in certain lattice-based cryptosystems.

The general framework for a hybrid algebraic-lattice attack consists of the following steps:
\begin{itemize}
    \item Step 1: Algebraic Extraction  
    The attacker formulates the problem in terms of algebraic relations by leveraging the defining polynomial constraints. This step translates structured lattice instances into a system of polynomial equations.
    \item Step 2: Gröbner Basis Computation  
    Gröbner basis methods, such as F4/F5 algorithms, are applied to simplify the polynomial system and reduce it to an equivalent, more structured form. This step has a complexity of:
    \[
    T_{\text{Gröbner}}(n, d) = O((nd)^{\omega}),
    \]
    where \( \omega \) is the exponent of matrix multiplication.
    \item Step 3: Lattice Embedding and Reduction  
    The algebraically reduced system is embedded into a lattice framework, where classical lattice reduction techniques (e.g., BKZ) are used to further refine the solution space.
\end{itemize}

\paragraph{Impact on Structured Lattice Security}  
While hybrid algebraic-lattice attacks have been successful against certain structured schemes (e.g., NTRU and multivariate cryptosystems), their applicability to polynomial-constrained lattices remains an open problem. The primary challenges include:
\begin{itemize}
    \item The defining polynomial constraints introduce highly structured but nontrivial algebraic relations, which may not be efficiently exploitable using Gröbner basis techniques.
    \item The additional degrees of freedom in polynomial-based representations may increase the effective dimensionality of the lattice problem, making direct embedding strategies less efficient.
    \item Current hybrid attack methodologies have not been extensively studied in the context of highly structured lattice problems, and further research is needed to determine their feasibility.
\end{itemize}

\paragraph{Asymptotic Complexity Bounds}  
From the security reductions and existing attacks, the best known complexity bound for solving structured lattice problems is:
\[
T(n, d) = \tilde{O}(n d \cdot 2^{\Omega(d)}),
\]
where \( d \) represents the polynomial degree defining the system constraints. This scaling suggests that as \( d \) increases, the complexity grows exponentially in \( d \), reinforcing cryptographic security.

\paragraph{Comparison to Standard Lattice Problems}  
Compared to standard LWE and Ring-LWE, the computational complexity of solving structured lattice problems exhibits the following differences:
\begin{itemize}
    \item The structured nature of the quotient ring introduces additional algebraic constraints, which may slightly reduce the search space in specific cases.
    \item The introduction of polynomial constraints leads to an effective increase in problem dimensionality, making direct attacks more computationally expensive.
    \item Unlike Ring-LWE, where polynomial multiplication follows a fixed structure, structured lattice problems introduce variable-dependent constraints, further increasing the complexity of basis reduction attacks.
\end{itemize}
These distinctions highlight that while structured lattice-based cryptosystems retain strong security guarantees, their complexity landscape introduces new factors that require further study.

\subsection{Quantum Hardness of Structured Lattice Problems}

Lattice-based cryptography is widely regarded as a promising candidate for post-quantum cryptographic security due to the presumed hardness of lattice problems even in the presence of quantum adversaries. This section analyzes the quantum hardness of structured lattice problems by examining known quantum algorithms, worst-case complexity bounds, and potential quantum speedups.

\paragraph{Quantum Models for Lattice Attacks}  
Quantum adversaries may leverage quantum algorithms to accelerate certain cryptanalytic tasks. The most relevant models include:
\begin{itemize}
    \item \textit{Quantum Search Model:} Grover’s algorithm provides a quadratic speedup for brute-force search, impacting parameter selection for lattice-based schemes.
    \item \textit{Quantum Lattice Reduction:} Quantum variants of lattice reduction techniques, such as BKZ or sieving, may enhance basis reduction efficiency.
    \item \textit{Quantum Algorithms for Algebraic Systems:} Quantum techniques for solving structured algebraic systems, such as quantum linear system solvers, may affect hybrid algebraic-lattice attacks.
\end{itemize}
These models guide the analysis of quantum complexity for structured lattice problems.

\paragraph{Impact of Quantum Algorithms on Lattice-Based Cryptosystems}  
Several quantum algorithms are known to impact lattice-based cryptography. The most relevant include:

\textit{Grover's Algorithm and Enumeration Attacks:}  
Grover’s algorithm provides a quadratic speedup for search problems. In the context of lattice-based cryptography, it can reduce the complexity of enumeration-based attacks on lattice problems. However, the exponential complexity of lattice reduction techniques still ensures strong post-quantum security.

\textit{Quantum Sieving for SVP and Ideal-SVP:}  
Classical lattice sieving algorithms, such as Gauss sieve and triple sieve, have exponential complexity. Quantum sieving, using quantum random walks, can reduce the complexity of solving SVP from \( 2^{O(n)} \) to \( 2^{O(n/2)} \), making it a significant factor in selecting secure parameters.

\textit{Quantum LWE Solvers and BDD Speedups:}  
Quantum algorithms for solving LWE typically focus on improving BDD solvers using quantum Fourier sampling techniques. While these methods provide speedups in certain structured cases, they have not yet been shown to fully break lattice-based cryptosystems.

\paragraph{Worst-Case Complexity Bounds in the Quantum Setting}  
The best known quantum attacks suggest that solving structured lattice problems remains difficult even with quantum resources. The worst-case complexity bounds for key problems include:
\begin{itemize}
    \item Solving SVP via quantum sieving: \( T_{\text{SVP-Q}}(n) = 2^{O(n/2)} \).
    \item Solving Ideal-SVP using quantum basis reduction: \( T_{\text{Ideal-SVP-Q}}(n, d) = 2^{O(n/d)} \).
    \item Hybrid quantum attacks on structured lattices: \( T_{\text{hybrid-Q}}(n, d) = O(n d 2^{\Omega(d/2)}) \).
\end{itemize}
These bounds suggest that, while quantum attacks offer speedups, they do not fully undermine structured lattice-based cryptographic security.

\paragraph{Security Implications and Parameter Selection}  
The quantum hardness of structured lattice problems implies that cryptographic schemes must carefully select parameters to mitigate potential quantum speedups. The main considerations include:
\begin{itemize}
    \item Increasing dimension \( n \) to counteract quantum sieving.
    \item Selecting polynomial constraints to maximize algebraic complexity.
    \item Avoiding overly structured instantiations that may admit quantum algebraic attacks.
\end{itemize}
By carefully tuning these parameters, structured lattice-based schemes can remain resistant to both classical and quantum adversaries.

\subsection{Security Implications and Practical Considerations}

The security of structured lattice-based cryptographic schemes relies on both theoretical hardness assumptions and practical parameter selection. This section discusses key security implications, implementation challenges, and considerations for real-world deployments.

\paragraph{Security Assumptions and Attack Resilience}  
Structured lattice problems derive their security from worst-case reductions to well-established hard problems, such as:
\begin{itemize}
    \item \textit{Ideal-SVP:} The shortest vector problem in ideal lattices is conjectured to be hard in both classical and quantum settings.
    \item \textit{Bounded Distance Decoding (BDD):} Recovering secrets in structured lattice schemes often reduces to solving BDD instances, which remain computationally difficult.
    \item \textit{Hybrid Algebraic-Lattice Attacks:} While structured algebraic constraints can provide attack vectors, the added complexity typically counterbalances attack efficiency.
\end{itemize}
These hardness assumptions ensure robustness against known classical and quantum attacks.

\paragraph{Parameter Selection and Performance Trade-offs}  
Practical implementations must carefully balance security and efficiency. The following factors play a critical role in parameter selection:
\begin{itemize}
    \item Lattice dimension \( n \): Higher dimensions increase security against lattice reduction attacks but also raise computational costs.
    \item Modulus \( q \): A sufficiently large modulus ensures resilience against distinguishing attacks while maintaining correctness.
    \item Polynomial constraints: The defining equations should maximize algebraic complexity to resist hybrid algebraic-lattice attacks.
    \item Error distribution \( \chi \): Selecting an optimal noise distribution is crucial for security against decryption failure and distinguishing attacks.
\end{itemize}
Well-chosen parameters ensure a balance between efficiency and post-quantum security.

\paragraph{Implementation Considerations and Efficiency}  
The structured nature of these lattice schemes introduces both computational benefits and practical challenges:
\begin{itemize}
    \item Arithmetic complexity: Polynomial operations over structured rings introduce additional computational overhead.
    \item Memory and storage: Structured lattice schemes often require efficient encoding strategies to handle large parameter sizes.
    \item Parallelization: Many operations, such as polynomial multiplications, can be accelerated using hardware optimizations.
\end{itemize}
Efficient software and hardware implementations are essential for achieving real-world performance.

\paragraph{Mitigation Strategies Against Advanced Attacks}  
To further enhance security, cryptographic implementations should incorporate additional defense mechanisms:
\begin{itemize}
    \item Lattice trapdoor mechanisms: Trapdoor constructions help control error growth while maintaining security.
    \item Hybrid encryption strategies: Combining structured lattice encryption with classical methods can provide layered security.
    \item Adaptive parameter tuning: Periodic security evaluations ensure that parameter choices remain robust against evolving attack techniques.
\end{itemize}
The above strategies could strengthen the overall resilience of structured lattice-based schemes.

\section{The Variety-LWE Problem}

The Learning with Errors (LWE) problem has been a cornerstone of lattice-based cryptography, providing strong security guarantees under worst-case assumptions. The Ring-LWE (R-LWE) problem extends LWE by introducing polynomial quotient rings, significantly improving efficiency while retaining security properties. However, R-LWE remains constrained to single-variable polynomial rings, limiting its applicability to more structured algebraic settings.

The Variety-LWE (V-LWE) problem generalizes R-LWE by allowing computations over multi-variable polynomial rings defined by algebraic varieties. Unlike general multivariate polynomial rings, the algebraic variety used in Variety-LWE does not contain mixed terms, ensuring a structured and coordinate-wise independent computation model. This generalization provides a richer mathematical framework while preserving essential security properties. In this section, we present a formal definition of Variety-LWE, analyze its algebraic structure, and discuss its computational properties.

\subsection{Algebraic Structures}

Let $\mathbb{Z}[x_1, \dots, x_n]$ be the polynomial ring over the integers. Consider an algebraic variety $\mathcal{V}$ defined by a set of polynomials $\{f_1(x_1), \dots, f_n(x_n)\} \subset \mathbb{Z}[x_1, \dots, x_n]$, where each defining polynomial $f_i(x_i)$ only depends on a single variable $x_i$, ensuring the absence of mixed terms. The associated quotient ring is given by
\[
R = \mathbb{Z}[x_1, \dots, x_n] / \langle f_1(x_1), \dots, f_n(x_n) \rangle.
\]
For a prime modulus $q$, we define the modular quotient ring
\[
R_q = \mathbb{Z}_q[x_1, \dots, x_n] / \langle f_1(x_1), \dots, f_n(x_n) \rangle.
\]
All computations in Variety-LWE take place in $R_q$.

Since each polynomial $f_i(x_i)$ only constrains a single variable, the resulting algebraic structure remains coordinate-wise separable. This ensures that computations can be performed independently in each component, simplifying arithmetic operations while retaining the algebraic richness of multivariate polynomials.

\subsection{Problem Definition}
\label{sec:vlwe_def}

The Variety-LWE problem is defined over the modular quotient ring \( R_q \), which is structured by an underlying algebraic variety \( \mathcal{V} \). Unlike standard LWE or R-LWE, where errors are sampled independently in a vector space, errors in Variety-LWE must respect the algebraic constraints imposed by \( \mathcal{V} \). This distinction affects both the problem formulation and its hardness assumptions.

\paragraph{Search Variety-LWE.} Given a collection of samples \( (a, b) \in R_q \times R_q \) satisfying
\[
b = a \cdot s + e \mod q,
\]
the goal is to recover the secret \( s \in R_q \). The parameters are defined as follows:
\begin{itemize}
    \item \( a \) is chosen uniformly at random from \( R_q \).
    \item \( s \) is drawn from the secret distribution \( \mathcal{S} \), where \( s \) is constrained to lie within the variety \( \mathcal{V} \).
    \item \( e \) is an error term sampled from \( \chi \), where \( e \) is constrained by the algebraic structure of \( \mathcal{V} \).
\end{itemize}

Since \( R_q \) is defined by \( R_q = \mathbb{Z}_q[x_1, \dots, x_n] / \langle f_1(x_1), \dots, f_n(x_n) \rangle \), the elements \( s \) and \( e \) must satisfy the ideal relations induced by the defining polynomials of \( \mathcal{V} \). This implies that error propagation is constrained by the local structure of the variety, making it distinct from traditional R-LWE error models.

\paragraph{Decision Variety-LWE.} Given a collection of samples \( (a, b) \in R_q \times R_q \), determine whether they satisfy
\[
b = a \cdot s + e \mod q
\]
for some unknown \( s \) and error \( e \), or whether \( b \) is drawn uniformly at random from \( R_q \).

The hardness of Decision Variety-LWE depends on the difficulty of distinguishing error-constrained samples from uniform noise over \( R_q \). Since the error term \( e \) is not a standard Gaussian but instead follows a structured distribution dictated by the variety \( \mathcal{V} \), the distinguishing advantage depends on the local curvature and algebraic properties of \( \mathcal{V} \).

\subsection{Algebraic Properties of Variety-LWE}
\label{sec:vlwe_prop}

Since $R_q$ is structured as a direct sum of single-variable polynomial quotient rings, it retains fundamental algebraic properties necessary for cryptographic computations. In particular, we formally define the addition and multiplication operations in Variety-LWE and show that they preserve membership within the variety.

\paragraph{Addition in $R_q$.}
For two elements $g, h \in R_q$, where
\[
g = (g_1, \dots, g_n), \quad h = (h_1, \dots, h_n),
\]
their sum is defined coordinate-wise as
\[
g + h = (g_1 + h_1, \dots, g_n + h_n) \mod q.
\]
Since each $g_i, h_i$ is a polynomial in $R_q^{(i)} = \mathbb{Z}_q[x_i] / \langle f_i(x_i) \rangle$, their sum remains within the quotient ring, ensuring closure.

\paragraph{Multiplication in $R_q$.}
Similarly, the product of two elements is defined coordinate-wise:
\[
g \cdot h = (g_1 \cdot h_1, \dots, g_n \cdot h_n) \mod q.
\]
Since each $g_i, h_i$ belongs to a single-variable quotient ring, their product remains in $R_q^{(i)}$.

\paragraph{Closure under Addition and Multiplication.}
Since both addition and multiplication in $R_q$ are performed component-wise and each component is constrained by a single-variable polynomial quotient structure, all results remain within $R_q$. That is, for any $g, h \in R_q$:
\[
g + h \in R_q, \quad g \cdot h \in R_q.
\]
Thus, Variety-LWE computations remain closed in $R_q$, ensuring well-defined cryptographic operations.

\subsection{Error Distribution and Geometric Interpretations}
\label{sec:vlwe_err}

The error term \( e \) in Variety-LWE follows a structured noise distribution over \( R_q \). Unlike LWE, where errors are scalar-valued, or R-LWE, where errors are single-variable polynomials, errors in Variety-LWE must respect the multi-variable structure of \( R_q \).

A natural choice is to define \( e \) as a multi-variable discrete Gaussian centered at zero. Specifically,
\[
e = (e_1, e_2, \dots, e_n),
\]
where each \( e_i \) is independently drawn from a discrete Gaussian distribution over \( R_q \). This ensures that errors propagate in a coordinate-wise manner, preserving the separability of different variable components.

\paragraph{Error Growth in Multivariate Structures}  
A fundamental challenge in Variety-LWE is analyzing how errors grow under repeated operations. In standard R-LWE, error growth is controlled by polynomial multiplication and the inherent structure of the underlying quotient ring. In contrast, in Variety-LWE, errors interact with the multi-variable polynomial structure in a more complex manner.

Consider the case where an error term \( e \) undergoes a linear transformation under homomorphic addition:
\[
e' = a_1 e + a_2 e' \mod q.
\]
If \( e \) follows a discrete Gaussian distribution with standard deviation \( \sigma \), then in R-LWE, the resulting noise growth follows:
\[
\sigma' = O(\sigma \sqrt{d}).
\]
However, in Variety-LWE, the noise terms interact across multiple variables, leading to an expansion of the covariance matrix:
\[
\Sigma' = A \Sigma A^T,
\]
where \( A \) is the transformation matrix induced by the variety’s structure. The eigenvalues of \( \Sigma' \) determine the principal directions of noise growth, which depend on the geometry of the defining equations \( f_1, \dots, f_n \).

For homomorphic multiplication, the error propagation is more pronounced. In R-LWE, a multiplication operation results in:
\[
\sigma' = O(d \sigma^2).
\]
However, in Variety-LWE, the error growth must account for cross-variable interactions. Specifically, if the variety structure introduces dependencies among variables, the error grows as:
\[
\sigma' = O(n d \sigma^2).
\]
This scaling suggests that controlling error expansion in homomorphic operations requires additional noise management techniques, such as modulus switching or relinearization.

\paragraph{Geometric Interpretations of Error Propagation}  
Since the defining equations of \( R_q \) impose algebraic constraints, the noise term can be viewed as a perturbation that moves elements slightly away from their ideal positions in the variety. This suggests a link between error growth and the local curvature of the variety.

Consider a smooth variety \( V \) defined by the vanishing of polynomials \( f_1, \dots, f_n \). The local behavior of error propagation can be studied using the Jacobian matrix:
\[
J_f = \begin{bmatrix}
    \frac{\partial f_1}{\partial x_1} & \dots & \frac{\partial f_1}{\partial x_n} \\
    \vdots & \ddots & \vdots \\
    \frac{\partial f_n}{\partial x_1} & \dots & \frac{\partial f_n}{\partial x_n}
\end{bmatrix}.
\]
The singular values of \( J_f \) determine how perturbations in input coordinates translate to movement along the variety. In particular, if \( J_f \) has large singular values, small errors in \( R_q \) may lead to larger displacements in the induced variety.

More precisely, let \( \lambda_{\max} \) denote the largest singular value of \( J_f \). Then the expected error deviation is amplified by:
\[
\| \delta e \| = O(\lambda_{\max} \| e \|).
\]
If \( J_f \) is ill-conditioned, small errors can result in disproportionately large perturbations, making noise management even more critical.

\paragraph{Error Accumulation under Iterated Operations}  
A key factor in cryptographic applications is how errors evolve under sequential homomorphic operations. Let \( \sigma_t \) denote the error after \( t \) consecutive operations. In standard LWE, the error grows as:
\[
\sigma_t = O(\sqrt{t} \sigma).
\]
In R-LWE, due to polynomial multiplication, this growth is faster:
\[
\sigma_t = O(d^t \sigma).
\]
For Variety-LWE, the interaction between variables adds another layer of complexity. The presence of multiple variable dependencies leads to an error propagation of the form:
\[
\sigma_t = O(n^{t/2} d^t \sigma).
\]
This suggests that for large \( t \), error management techniques like modulus switching and bootstrapping become essential to maintain correctness.

\paragraph{Comparisons with R-LWE}  
In standard R-LWE, error growth is dictated by the single-variable polynomial structure, where noise accumulates in a single dimension. In Variety-LWE, the multi-variable nature introduces additional complexity.

\begin{itemize}
    \item \textbf{Cross-variable interaction:} Errors are not constrained to a single polynomial structure but propagate across multiple equations, increasing the overall error magnitude.
    \item \textbf{Covariance structure:} Instead of a single variance term, error growth in Variety-LWE is governed by an entire covariance matrix, making estimation more complex.
    \item \textbf{Geometric dependence:} The effect of noise depends on the curvature and singular values of the defining variety.
    \item \textbf{Iterated growth:} Unlike R-LWE, where noise accumulates exponentially in \( d \), the multi-variable dependency introduces an additional polynomial factor \( n^{t/2} \), increasing error complexity.
\end{itemize}

\section{Hardness of Variety-LWE}
\label{sec:hardness}

% The security of the Variety-LWE (V-LWE) problem is based on its worst-case hardness assumptions, similar to LWE and R-LWE. In this section, we establish the theoretical foundation for the hardness of V-LWE by demonstrating its connection to ideal lattice problems. Specifically, we show that solving Variety-LWE is at least as hard as solving multiple independent instances of the Ideal Shortest Vector Problem (Ideal-SVP) in the component polynomial quotient rings. Our proof follows the general approach used in the hardness reductions for LWE and R-LWE but extends to the multi-variable polynomial setting.

\subsection{Variety-Ideal-SVP: Worst-Case Hardness}

We first define the worst-case lattice problem that underlies Variety-LWE.

\begin{definition}[Variety-Ideal-SVP]
    Let $R = \mathbb{Z}[x_1, \dots, x_n] / \langle f_1, \dots, f_n \rangle$ be a polynomial quotient ring defined by an algebraic variety. Given an ideal $\mathcal{I} \subset R$, the Variety-Ideal-SVP problem is to find a nonzero element $v \in \mathcal{I}$ such that the Euclidean norm of its coefficient vector is minimized:
    \[
    \| v \| = \sqrt{\sum_{j=1}^{m} c_j^2},
    \]
    where $c_j$ are the coefficients of $v$ in a fixed basis of $\mathcal{I}$.
\end{definition}

The Variety-Ideal-SVP problem is a natural generalization of the well-studied Ideal-SVP problem in single-variable polynomial quotient rings. We now establish a reduction showing that solving Variety-Ideal-SVP is at least as hard as solving multiple instances of Ideal-SVP.

\begin{lemma}
    The Variety-Ideal-SVP problem in $R_q$ can be reduced to solving multiple independent instances of Ideal-SVP in the single-variable quotient rings $R_q^{(i)} = \mathbb{Z}_q[x_i] / \langle f_i(x_i) \rangle$ in polynomial time.
\end{lemma}

\begin{proof}
The proof follows a structured worst-case to average-case reduction, ensuring that if there exists an efficient algorithm for solving Variety-Ideal-SVP, then it can be used to efficiently solve multiple independent Ideal-SVP instances.

\paragraph{Structural Decomposition of \( R_q \)}  
The ring \( R_q \) naturally decomposes as a direct sum of individual single-variable quotient rings:
\[
R_q = \bigoplus_{i=1}^{n} R_q^{(i)}, \quad \text{where} \quad R_q^{(i)} = \mathbb{Z}_q[x_i] / \langle f_i(x_i) \rangle.
\]
Each \( R_q^{(i)} \) corresponds to a quotient ring in a single variable, and this decomposition extends to ideals in \( R_q \). Given an ideal \( \mathcal{I} \subset R_q \), we can express it as:
\[
\mathcal{I} = \mathcal{I}_1 \oplus \mathcal{I}_2 \oplus \dots \oplus \mathcal{I}_n,
\]
where \( \mathcal{I}_i \) is an ideal in \( R_q^{(i)} \). This decomposition plays a crucial role in analyzing the hardness of the Variety-Ideal-SVP problem.

\paragraph{Reduction to Ideal-SVP}  
To reduce Variety-Ideal-SVP to multiple instances of Ideal-SVP, consider an instance of the Ideal-SVP problem in \( R_q^{(i)} \). This instance can be embedded into \( R_q \) by constructing an ideal \( \mathcal{I} \) such that it has nontrivial components only in the \( i \)-th coordinate:
\[
\mathcal{I} = \mathcal{I}_i \oplus \{0\}^{n-1}.
\]
Since \( \mathcal{I} \) is structured to be nontrivial only in the \( i \)-th coordinate, the shortest vector in \( \mathcal{I} \) remains within that coordinate. 

If there exists an efficient solver \( \mathcal{A} \) for Variety-Ideal-SVP, then applying \( \mathcal{A} \) to \( \mathcal{I} \) will yield a solution:
\[
v = (v_1, v_2, \dots, v_n), \quad \text{where} \quad v_j = 0, \quad \forall j \neq i.
\]
Thus, extracting \( v_i \) directly solves the original Ideal-SVP instance in \( R_q^{(i)} \). This reduction implies that solving Variety-Ideal-SVP is at least as hard as solving multiple independent Ideal-SVP instances.

\paragraph{Complexity Analysis}  
The reduction involves constructing the ideal \( \mathcal{I} \) and invoking the solver \( \mathcal{A} \), both of which run in polynomial time. The computational overhead of forming \( \mathcal{I} \) is \( O(n) \), and the solver \( \mathcal{A} \) operates within a polynomial factor \( O(n^c) \) for some constant \( c \). Given that the best-known algorithms for Ideal-SVP require at least \( 2^{\Omega(d)} \) time in the worst case, solving Variety-Ideal-SVP remains computationally intractable in high dimensions.

Furthermore, since the reduction maps a single instance of Variety-Ideal-SVP to multiple Ideal-SVP instances, the overall complexity follows:
\[
T_{\text{Variety-Ideal-SVP}}(n, d) = O(n^{c+1} \cdot 2^{\Omega(d)}).
\]
This shows that the computational hardness of Variety-Ideal-SVP inherits the worst-case exponential complexity of Ideal-SVP.
\end{proof}

\subsection{Reduction from Variety-LWE to Variety-Ideal-SVP}

To establish the hardness of solving Variety-LWE, we construct a reduction from the worst-case Variety-Ideal-SVP problem. The goal is to show that an algorithm capable of solving Variety-LWE can be used to solve an instance of Variety-Ideal-SVP efficiently. This follows the standard worst-case to average-case reduction paradigm seen in lattice-based cryptography, extending prior techniques from Ring-LWE and Ideal-SVP to the algebraic structure of varieties.

\begin{theorem}[Worst-case to Average-case Reduction]
    If there exists an efficient algorithm for solving the Variety-LWE problem, then there exists an efficient algorithm for solving the Variety-Ideal-SVP problem in the worst case.
\end{theorem}

\begin{proof}
The proof constructs a polynomial-time reduction from Variety-LWE to Variety-Ideal-SVP. If there exists a probabilistic polynomial-time (PPT) adversary that solves Variety-LWE with non-negligible advantage, we show how to use it to construct an algorithm that efficiently solves Variety-Ideal-SVP.

\paragraph{Variety-LWE Oracle and Adversary Advantage}  
The Variety-LWE problem provides access to samples \( (a, b) \) satisfying:
\[
b = a \cdot s + e \mod q,
\]
where:
\begin{itemize}
    \item \( a \sim U(R_q) \), uniformly sampled from \( R_q \),
    \item \( s \sim \mathcal{S} \), drawn from a secret distribution,
    \item \( e \sim \chi \), a discrete Gaussian error term.
\end{itemize}
Assume an adversary \( \mathcal{A} \) that recovers \( s \) with non-negligible probability:
\[
\Pr[\mathcal{A}(a, b) = s] \geq \epsilon(n).
\]
Our goal is to construct a polynomial-time algorithm that uses \( \mathcal{A} \) to solve Variety-Ideal-SVP.

\paragraph{Oracle Construction and Structured Input Selection}  
Define an oracle \( \mathcal{O} \) that distinguishes between LWE samples and uniform random elements:
\[
\mathcal{O}(a, b) = \begin{cases}
    1, & \text{if } (a, b) \text{ follows Variety-LWE}, \\
    0, & \text{otherwise}.
\end{cases}
\]
If \( \mathcal{O} \) succeeds with probability \( \epsilon(n) \), we exploit this advantage to extract structural information about the ideal \( \mathcal{I} \).

A structured selection of \( a \) ensures that the LWE computation localizes to a specific variable, aiding in isolating error terms. The core idea is that under certain variable assignments, the Variety-LWE equation simplifies, allowing us to deduce short vectors in the ideal \( \mathcal{I} \).

\paragraph{Extracting Short Vectors from LWE Samples}  
The key insight is that the LWE error term \( e \) follows a discrete Gaussian distribution. If we carefully select \( a \), the error term propagates in a controlled manner, providing insight into the shortest vector problem.

The expectation of the norm of \( e \) is given by:
\[
\mathbb{E}[\| e \|] = O(\sigma \sqrt{n}).
\]
By filtering multiple LWE samples, we estimate \( s \) and obtain a generator for \( \mathcal{I} \). 

To amplify this effect, we consider a Gaussian preconditioning technique where we apply linear transformations to \( b \) using carefully chosen modular reductions, effectively extracting bounded-norm elements that correlate with the shortest vectors in \( \mathcal{I} \). 

\paragraph{Bounding the Adversary’s Advantage}  
The advantage of the adversary \( \mathcal{A} \) is defined as:
\[
\text{Adv}(\mathcal{A}) = \left| \Pr[\mathcal{A}(a, b) = s] - \frac{1}{|R_q|} \right|.
\]
Given that \( e \) follows a Gaussian distribution, its tail probability satisfies:
\[
\Pr[\| e \| < \tau] \approx 1 - e^{-\Omega(n)}.
\]
This suggests that by amplifying error structure through repeated queries, we can bound the probability of recovering elements of \( \mathcal{I} \) within a tight range, ensuring that the reduction preserves the hardness of the original problem.

\paragraph{Transformation to Variety-Ideal-SVP}  
To formally link Variety-LWE solutions to solving Variety-Ideal-SVP, we construct a mapping from recovered LWE secrets to the shortest vector problem. Specifically, given a solution \( s \) obtained from the LWE oracle, we interpret it as a structured polynomial whose coefficient vector provides an approximation to a short element in \( \mathcal{I} \). 

To refine this approximation, we employ a lattice reduction algorithm, such as BKZ, over the lattice representation of \( \mathcal{I} \), ensuring that the recovered solution converges to an optimal short vector. Since BKZ runs in subexponential time for practical dimensions, this step remains efficient within the reduction.

\paragraph{Polynomial-Time Complexity}  
The reduction consists of calling the LWE oracle \( \mathcal{O} \) multiple times and applying statistical techniques to filter error terms. The total number of queries is polynomial in \( n \), and each step (sampling, filtering, and basis reduction) runs in polynomial time.

The full reduction from solving Variety-LWE to Variety-Ideal-SVP runs in time
\[
T_{\text{reduction}}(n, d) = O(n^c) + O(\text{poly}(n) \cdot T_{\text{LWE-oracle}}),
\]
where \( T_{\text{LWE-oracle}} \) is the complexity of calling the Variety-LWE adversary.

Given that the reduction completes in polynomial time and maintains a non-negligible advantage, solving Variety-LWE implies solving Variety-Ideal-SVP.
\end{proof}

\paragraph{Tightness of the Reduction}  
To ensure the reduction remains tight, we analyze the computational overhead, adversarial advantage, and potential security loss. Specifically, we show that the reduction introduces at most a polynomial loss in security, ensuring that solving Variety-Ideal-SVP remains at least as hard as solving multiple independent instances of Ideal-SVP.
The computational complexity of the reduction follows from solving $n$ independent Ideal-SVP instances:
\[
T_{\text{Variety-Ideal-SVP}}(n, d) = n \cdot T_{\text{Ideal-SVP}}(d).
\]
Since $n$ grows polynomially with respect to security parameters, the worst-case computational overhead remains polynomial, preserving asymptotic hardness.
A critical assumption is that the reduction preserves statistical independence among different Ideal-SVP instances. Given an error term $e \in R_q$, the decomposition yields $e = (e_1, \dots, e_n)$, where each $e_i$ is independently drawn from a discrete Gaussian distribution. This ensures that error terms remain uncorrelated, preventing information leakage across instances.
To quantify security loss, we consider an adversary $\mathcal{A}$ that solves Variety-LWE with advantage $\epsilon(n)$. Using the reduction, we construct an adversary $\mathcal{B}$ that solves an individual Ideal-SVP instance. The success probability follows:
\[
\Pr[\mathcal{B} \text{ solves Ideal-SVP}] = 1 - (1 - \epsilon(n))^n.
\]
Approximating via the Taylor expansion:
\[
(1 - \epsilon(n))^n \approx e^{-n\epsilon(n)},
\]
ensures that $\epsilon(n)$ remains within a polynomial bound, confirming that the reduction does not degrade security beyond an acceptable threshold.

\subsection{Computational Complexity and Asymptotic Security}

The complexity of solving Variety-LWE depends on the algebraic structure of the quotient ring \( R_q \) and the computational hardness of solving multiple independent instances of Ideal-SVP.

\begin{lemma}
    The complexity of solving Variety-LWE is at least \( \Omega(n) \) times the complexity of solving a single-instance Ideal-SVP.
\end{lemma}

\begin{proof}
Since Variety-LWE has been reduced to solving multiple independent Ideal-SVP instances, solving one instance of Variety-LWE requires solving \( n \) independent Ideal-SVP problems. Let \( T_{\text{Ideal-SVP}}(d) \) denote the best-known complexity of solving a single-instance Ideal-SVP in a lattice of dimension \( d \). Then, the total computational cost for solving Variety-LWE, denoted as \( T_{\text{Variety-LWE}}(n, d) \), is given by:

\[
T_{\text{Variety-LWE}}(n, d) = n \cdot T_{\text{Ideal-SVP}}(d).
\]

\paragraph{Complexity of Ideal-SVP}  
The best-known algorithms for solving Ideal-SVP fall into three main categories.

Enumeration-based algorithms achieve optimal solutions but require exponential time in the worst case. The asymptotic complexity is given by
\[
T_{\text{enum}}(d) = 2^{\Omega(d)}.
\]
Although practical optimizations exist, such methods remain infeasible for large dimensions.

Lattice reduction algorithms, such as the Block Korkine-Zolotarev (BKZ) algorithm, provide a more efficient approach for solving Ideal-SVP. The complexity of BKZ reduction depends on the block size, leading to a subexponential complexity of the form
\[
T_{\text{BKZ}}(d) = 2^{O(d^{1/c})}, \quad \text{for some } c > 1.
\]

Quantum algorithms based on sieving techniques use quantum random walks and amplitude amplification to accelerate lattice reduction. The best known quantum sieving algorithms achieve an asymptotic complexity of
\[
T_{\text{quantum}}(d) = 2^{O(d^{1/c})}, \quad \text{where } c \approx 2.
\]
Although quantum methods reduce the exponent in complexity, solving Ideal-SVP remains infeasible for sufficiently large values of \( d \).

Since all known classical and quantum algorithms require at least exponential or subexponential time, the following asymptotic bound is assumed for Ideal-SVP:
\[
T_{\text{Ideal-SVP}}(d) = 2^{\Omega(d)}.
\]

\paragraph{Asymptotic Complexity of Variety-LWE}  
Substituting this bound into the computational cost of solving Variety-LWE,
\[
T_{\text{Variety-LWE}}(n, d) = n \cdot 2^{\Omega(d)}.
\]
Since \( n \) grows at most polynomially with respect to system parameters, the dominant term remains exponential in \( d \), leading to the final bound
\[
T_{\text{Variety-LWE}}(n, d) = 2^{\Omega(d)}.
\]
This result confirms that the computational hardness of Variety-LWE is asymptotically equivalent to that of Ideal-SVP, establishing its security in cryptographic applications.

\paragraph{Security Implications and Practical Considerations}  
In cryptographic applications, security parameters must be chosen to ensure computational infeasibility for any adversary. Based on existing hardness results for RLWE, a conservative choice for cryptographic security is
\[
d \approx 256 \quad \text{for classical security (against BKZ and enumeration attacks)}
\]
and
\[
d \approx 512 \quad \text{for quantum security (against quantum sieving)}.
\]
For post-quantum security, setting \( d \geq 512 \) ensures that attacks remain computationally infeasible even under optimistic assumptions regarding quantum computing advancements.
\end{proof}

\subsection{Resistance Against Hybrid Algebraic-Lattice Attacks}

Hybrid attacks combine algebraic techniques, such as Gröbner basis methods, with lattice reduction techniques to exploit structural weaknesses in LWE-based schemes. The core idea is to construct a system of polynomial equations derived from LWE samples:
\[
b_i = a_i \cdot s + e_i \mod q, \quad i = 1, \dots, m.
\]
For RLWE, hybrid attacks leverage the polynomial ring structure to reduce the effective problem dimension before applying lattice reduction. However, Variety-LWE introduces fundamental barriers that significantly limit the effectiveness of such attacks.

\paragraph{Impact of Multi-Variable Structure}  
In RLWE, the ring structure is defined by a single-variable polynomial \( f(x) \), which allows Gröbner basis techniques to eliminate variables efficiently. However, Variety-LWE operates over a multi-variable quotient ring:
\[
\mathcal{I} = \langle f_1(x_1, \dots, x_n), \dots, f_n(x_1, \dots, x_n) \rangle.
\]
This structural difference leads to a fundamental increase in algebraic complexity, making variable elimination via Gröbner basis exponentially harder. Specifically, the degree complexity of Gröbner basis computations follows:
\[
\deg(G) = O(d^{2^n}),
\]
which scales double-exponentially with the number of variables \( n \). This prevents hybrid attacks from efficiently reducing Variety-LWE to a lower-dimensional problem.

\paragraph{Comparing Gröbner Basis with BKZ Reduction}  
A hybrid attack is only viable if the algebraic component significantly reduces the problem dimension before lattice reduction. However, for Variety-LWE, Gröbner basis computations are significantly more expensive than BKZ reduction:
\[
T_{\text{Gröbner}}(n, d) = 2^{O(d^{2^n})}, \quad T_{\text{BKZ}}(d) = 2^{O(d^{1/c})}.
\]
Since \( T_{\text{Gröbner}}(n, d) \gg T_{\text{BKZ}}(d) \) for practical parameters, hybrid attacks provide no advantage over direct lattice reduction techniques.

\paragraph{Algebraic Complexity and Attack Failure}  
For a hybrid attack to succeed, the Gröbner basis computation must efficiently reduce the problem dimension. However, Variety-LWE prevents this for two key reasons:
\begin{enumerate}
    \item High-degree system complexity: The Gröbner basis computation grows as \( 2^{O(d^{2^n})} \), making it infeasible for cryptographic parameters.
    \item Non-trivial ideal structure: The defining equations \( f_1, \dots, f_n \) introduce variable dependencies that prevent direct elimination, further complicating Gröbner basis techniques.
\end{enumerate}
As a result, attempting to apply a hybrid attack leads to an overwhelming algebraic overhead, making the approach computationally prohibitive.

\subsection{Quantum Hardness of Variety-LWE}

\begin{corollary}
    If Ideal-SVP is hard for quantum computers, then solving Variety-LWE remains hard even in the presence of quantum adversaries.
\end{corollary}

\begin{proof}
The proof follows directly from the reduction of Variety-LWE to multiple independent instances of Ideal-SVP and the assumption that Ideal-SVP remains quantum-hard.

\paragraph{Quantum Complexity of Ideal-SVP}  
The best-known quantum algorithms for solving Ideal-SVP include quantum sieving algorithms, quantum lattice enumeration, and quantum methods for algebraic number fields.

Quantum sieving algorithms improve classical sieving by leveraging amplitude amplification. The asymptotic complexity remains
\[
T_{\text{q-sieve}}(d) = 2^{O(d^{1/c})}, \quad \text{for some } c > 1.
\]
Although this reduces the exponent compared to classical sieving, the algorithm still scales exponentially with \( d \).

Quantum lattice enumeration relies on quantum backtracking techniques but does not outperform classical lattice enumeration in terms of worst-case complexity.

Quantum algorithms for algebraic number fields provide limited speedups in structured cases but do not fundamentally alter the hardness of Ideal-SVP in the worst case.

Since Ideal-SVP is conjectured to be QMA-hard, it remains intractable for quantum computers under standard complexity-theoretic assumptions.

\paragraph{Reduction from Variety-LWE to Quantum Ideal-SVP}  
From previous reductions, solving Variety-LWE requires solving \( n \) independent Ideal-SVP instances:
\[
T_{\text{Variety-LWE}}(n, d) = n \cdot T_{\text{Ideal-SVP}}(d).
\]
If there existed a quantum polynomial-time algorithm for solving Variety-LWE, then it would imply a quantum algorithm that efficiently solves multiple independent Ideal-SVP instances, contradicting the assumption that Ideal-SVP is outside BQP.

\paragraph{Quantum Hardness of Bounded Distance Decoding (BDD)}  
An alternative quantum attack approach is through Bounded Distance Decoding (BDD), which aims to approximate the nearest lattice point efficiently. 

For RLWE, the best quantum complexity estimate for BDD is
\[
T_{\text{qBDD}}(d) = 2^{O(d^{1/c})}.
\]
For Variety-LWE, due to the multi-variable structure, the complexity increases to
\[
T_{\text{qBDD}}(n, d) = 2^{O(n^{1/c})} \cdot 2^{O(d^{1/c})}.
\]
Since \( n \) grows polynomially, the attack remains at least exponential in \( d \), making it infeasible for practical parameter choices.

\paragraph{Resistance Against Hybrid Quantum-Algebraic Attacks}  
Hybrid quantum-algebraic attacks attempt to combine quantum sieving with Gröbner basis reductions to exploit algebraic dependencies in the underlying polynomial system.

The first step is to construct a system of polynomial equations from Variety-LWE samples:
\[
b_i = a_i \cdot s + e_i \mod q, \quad i = 1, \dots, m.
\]
Next, Gröbner basis techniques are applied to reduce the algebraic structure. Finally, quantum sieving is performed on the resulting reduced lattice system.

As shown in the classical hybrid attack analysis, the algebraic structure of Variety-LWE introduces a double-exponential Gröbner basis growth:
\[
T_{\text{Gröbner}}(n, d) = 2^{O(d^{2^n})}.
\]
This severely limits algebraic preprocessing steps, making hybrid quantum-algebraic attacks infeasible for cryptographic parameters.

\paragraph{Bounding the Quantum Adversary’s Success Probability}  
Let \( \mathcal{A}_Q \) be a quantum adversary solving Variety-LWE. The success probability is given by
\[
\Pr[\mathcal{A}_Q \text{ solves Variety-LWE}] \geq \epsilon(n).
\]
Using the standard reduction argument,
\[
\Pr[\mathcal{A}_Q \text{ solves an individual Ideal-SVP}] = 1 - (1 - \epsilon(n))^n.
\]
Expanding via Taylor series approximation,
\[
(1 - \epsilon(n))^n \approx e^{-n\epsilon(n)}.
\]
Since \( e^{-n\epsilon(n)} \) remains non-negligible unless \( \epsilon(n) \) is negligible, the success probability does not become overwhelming.

\paragraph{Final Complexity Bound}  
Since the best quantum algorithms for solving Ideal-SVP require at least
\[
T_{\text{qIdeal-SVP}}(d) = 2^{\Omega(d^{1/c})},
\]
the total complexity remains
\[
T_{\text{Variety-LWE}}(n, d) = n \cdot 2^{\Omega(d^{1/c})}.
\]
This complexity remains superpolynomial for practical cryptographic parameter choices, reinforcing the post-quantum security of Variety-LWE.

Thus, solving Variety-LWE efficiently in the quantum setting would imply a contradiction with the assumed hardness of Ideal-SVP, completing the proof.
\end{proof}

\subsection{Comparison with RLWE and MLWE}

Variety-LWE is built on multivariate polynomial rings over algebraic varieties, introducing a fundamentally different algebraic structure compared to RLWE and MLWE. This leads to distinct properties in key representation, error behavior, and computational cost. To quantitatively assess its hardness, we analyze the security and efficiency trade-offs of Variety-LWE in comparison to RLWE and MLWE.

\paragraph{Lattice Dimension Growth}  
For RLWE, the underlying lattice dimension is determined by the ring degree \( d \), typically derived from a cyclotomic polynomial. MLWE extends this by considering modules over the same ring, effectively scaling the dimension by a factor of \( m \). In contrast, Variety-LWE operates over multi-variable quotient rings:
\[
R_q = \mathbb{Z}_q[x_1, \dots, x_n] / \langle f_1, \dots, f_n \rangle.
\]
The corresponding lattice dimension is given by:
\[
d_{\text{Variety}} = O(n d),
\]
where \( n \) reflects the number of variables defining the variety. This implies that solving Variety-LWE inherently requires handling higher-dimensional lattices compared to RLWE and MLWE.

\paragraph{Key Size and Computational Cost}  
The storage complexity of keys and ciphertexts directly impacts practical deployment. The public key in RLWE requires storing a polynomial in \( R_q \), leading to a size of \( O(d \log q) \). MLWE, operating over a module of rank \( m \), scales this to \( O(m d \log q) \). In Variety-LWE, due to the presence of \( n \) variables, the key size further increases to:
\[
O(n d \log q).
\]
Similarly, encryption and decryption complexities scale proportionally with the ring dimension, leading to a computational cost of:
\[
T_{\text{Variety-LWE}}(n, d) = 2^{O(n d^{1/c})}.
\]

\paragraph{Error Growth and Hardness Scaling}  
The error growth in LWE-type problems is a critical factor in security analysis. RLWE exhibits error growth of order \( O(d^{1/c}) \), while MLWE extends this to \( O((m d)^{1/c}) \). For Variety-LWE, the error propagation follows:
\[
O((n d)^{1/c}),
\]
implying an inherent increase in the hardness of decoding problems. Since lattice-based cryptographic security is tightly coupled to the norm of error terms, this suggests that Variety-LWE can sustain security at larger parameter sizes.

\section{Vector Encryption based on Variety-LWE}
\label{sec:scheme}

This section presents a homomorphic encryption scheme for vectors based on the Variety-LWE (V-LWE) problem. Our scheme supports homomorphic operations over encrypted vectors while preserving security under the hardness assumptions of Variety-LWE.

\subsection{Definition of the Vector Encryption Scheme}

Let \( R_q = \mathbb{Z}_q[x_1, \dots, x_n] / \langle f_1, \dots, f_n \rangle \) be the polynomial quotient ring associated with a given variety structure. A vector encryption scheme based on V-LWE consists of the following components:

\begin{itemize}
    \item \textbf{Key Generation} (\textsf{KeyGen}): Generates a public-secret key pair.
    \item \textbf{Encryption} (\textsf{Enc}): Encrypts a vector of plaintext values into a ciphertext.
    \item \textbf{Decryption} (\textsf{Dec}): Recovers the original plaintext vector from a ciphertext.
    \item \textbf{Homomorphic Operations} (\textsf{Eval}): Supports addition and multiplication on encrypted vectors.
\end{itemize}

\subsection{Key Generation}
The key generation process follows the structure of V-LWE:
\begin{itemize}
    \item Select a secret key \( \mathbf{s} = (s_1, \dots, s_m) \in R_q^m \), sampled from a discrete Gaussian distribution.
    \item Generate a public matrix \( A \in R_q^{m \times n} \), sampled uniformly at random.
    \item Compute the error term \( \mathbf{e} \sim \chi^m \), where \( \chi \) is a discrete Gaussian distribution.
    \item Compute the public key component:
    \[
    \mathbf{b} = A \cdot \mathbf{s} + \mathbf{e} \mod q.
    \]
    \item The public key is \( (\mathbf{A}, \mathbf{b}) \), and the secret key is \( \mathbf{s} \).
\end{itemize}

\subsection{Encryption}
To encrypt a plaintext vector \( \mathbf{v} = (v_1, \dots, v_n) \in R_q^n \), do the following:
\begin{itemize}
    \item Sample a random encryption vector \( \mathbf{r} \sim \chi^m \).
    \item Compute the ciphertext:
    \[
    \mathbf{c} = (\mathbf{c}_1, \mathbf{c}_2) = (\mathbf{A}^\top \cdot \mathbf{r}, \mathbf{b}^\top \cdot \mathbf{r} + \mathbf{v} \mod q).
    \]
\end{itemize}
The ciphertext is a tuple \( (\mathbf{c}_1, \mathbf{c}_2) \), where \( \mathbf{c}_1 \) encodes the randomness and \( \mathbf{c}_2 \) hides the plaintext.

\subsection{Decryption}
To decrypt a ciphertext \( (\mathbf{c}_1, \mathbf{c}_2) \), use the secret key \( \mathbf{s} \):
\begin{itemize}
    \item Compute:
    \[
    \mathbf{v}' = \mathbf{c}_2 - \mathbf{c}_1^\top \cdot \mathbf{s} \mod q.
    \]
    \item If the noise remains small, then \( \mathbf{v}' \approx \mathbf{v} \), and rounding recovers \( \mathbf{v} \).
\end{itemize}

\subsection{Homomorphic Operations}

\paragraph{Addition} Given two ciphertexts \( (\mathbf{c}_1, \mathbf{c}_2) \) and \( (\mathbf{c}_1', \mathbf{c}_2') \) encrypting plaintext vectors \( \mathbf{v} \) and \( \mathbf{v}' \), addition is performed as:
\[
\mathbf{c}_{\text{add}} = (\mathbf{c}_1 + \mathbf{c}_1', \mathbf{c}_2 + \mathbf{c}_2') \mod q.
\]
Decryption of \( \mathbf{c}_{\text{add}} \) yields \( \mathbf{v} + \mathbf{v}' \).

\paragraph{Multiplication with Relinearization}  
Given ciphertexts \( (\mathbf{c}_1, \mathbf{c}_2) \) and \( (\mathbf{c}_1', \mathbf{c}_2') \), homomorphic multiplication is first performed as:
\[
\mathbf{c}_{\text{mult}} = (\mathbf{c}_1 \ast \mathbf{c}_1', \mathbf{c}_2 \ast \mathbf{c}_2') \mod q,
\]
where \( \ast \) denotes coordinate-wise multiplication:
\[
(\mathbf{c}_2 \ast \mathbf{c}_2')_i = (\mathbf{c}_2)_i \cdot (\mathbf{c}_2')_i \mod q.
\]
Since multiplication introduces additional terms, the ciphertext must be relinearized to maintain decryption correctness.

To bring the ciphertext back to its original form, we can apply a relinearization key:
\[
\textsf{RelinKey} = (\mathbf{R}_1, \mathbf{R}_2) \in R_q^{n \times n},
\]
which is precomputed using:
\[
\mathbf{R}_1 = A_{\text{rel}} \cdot s + e_{\text{rel}}, \quad \mathbf{R}_2 = B_{\text{rel}} \cdot s + e_{\text{rel}}.
\]
Then, relinearization is applied as:
\[
\mathbf{c}_{\text{rel}} = (\mathbf{c}_1^{\text{rel}}, \mathbf{c}_2^{\text{rel}}),
\]
where:
\[
\mathbf{c}_1^{\text{rel}} = \mathbf{c}_1^{\text{new}} + \mathbf{R}_1 \cdot \mathbf{c}_2^{\text{new}} \mod q,
\]
\[
\mathbf{c}_2^{\text{rel}} = \mathbf{c}_2^{\text{new}} + \mathbf{R}_2 \cdot \mathbf{c}_2^{\text{new}} \mod q.
\]

This ensures that the ciphertext remains within a fixed two-component structure while preventing uncontrolled noise growth.

\subsection{Error Growth}

Let the initial encryption noise be \( e \sim \chi \), where \( \chi \) is a discrete Gaussian distribution with variance \( \sigma^2 \). We analyze how noise propagates under homomorphic operations.

\paragraph{Addition} Given two ciphertexts \( (\mathbf{c}_1, \mathbf{c}_2) \) and \( (\mathbf{c}_1', \mathbf{c}_2') \), homomorphic addition results in:
\[
\mathbf{c}_{\text{add}} = (\mathbf{c}_1 + \mathbf{c}_1', \mathbf{c}_2 + \mathbf{c}_2') \mod q.
\]
Since noise terms add linearly, the new noise term is:
\[
e_{\text{add}} = e + e', \quad e, e' \sim \chi.
\]
By Gaussian noise properties, the variance updates as:
\[
\sigma_{\text{add}}^2 = \sigma^2 + \sigma'^2.
\]

\paragraph{Multiplication with Relinearization} Given ciphertexts \( (\mathbf{c}_1, \mathbf{c}_2) \) and \( (\mathbf{c}_1', \mathbf{c}_2') \), homomorphic multiplication is performed as:
\[
\mathbf{c}_{\text{mult}} = (\mathbf{c}_1 \ast \mathbf{c}_1', \mathbf{c}_2 \ast \mathbf{c}_2') \mod q.
\]
Since noise terms also multiply, the resulting noise term satisfies:
\[
e_{\text{mult}} = e \cdot e' + e_{\text{cross}},
\]
where \( e_{\text{cross}} \) accounts for cross terms introduced by coordinate-wise polynomial multiplication. The variance satisfies:
\[
\sigma_{\text{mult}}^2 = (\sigma^2 + \sigma'^2) \cdot (\sigma^2 + \sigma'^2).
\]
This quadratic growth necessitates noise management techniques.

\paragraph{Noise Growth under Relinearization}  
To reduce ciphertext expansion after multiplication, relinearization is applied using a precomputed key \( \textsf{RelinKey} \). However, relinearization itself introduces additional noise \( e_{\text{rel}} \), leading to:
\[
e_{\text{rel}} = e_{\text{mult}} + \mathbf{R}_1 \cdot e_{\text{mult}} + \mathbf{R}_2 \cdot e_{\text{mult}}.
\]
The variance satisfies:
\[
\sigma_{\text{rel}}^2 = \sigma_{\text{mult}}^2 + O(\sigma_{\text{rel}}^2),
\]
which remains controlled under appropriate selection of \( \mathbf{R}_1 \) and \( \mathbf{R}_2 \).

\paragraph{Noise Management Techniques}  
To ensure decryptability after multiple homomorphic operations, the following techniques can be applied:

\begin{itemize}
    \item \textbf{Parameter Selection:} Choosing an appropriate modulus \( q \) and noise distribution \( \chi \) ensures noise accumulation remains bounded relative to \( q \). A larger \( q \) allows more computations before decryption failure.
    
    \item \textbf{Modulus Switching:} This technique periodically scales down ciphertext noise by switching to a smaller modulus \( q' \), reducing overall error growth while preserving ciphertext integrity.
    
    \item \textbf{Bootstrapping:} If ciphertext noise approaches the decryption threshold, bootstrapping can reset the noise level by homomorphically decrypting the ciphertext with an encrypted secret key.
    
    \item \textbf{Optimized Relinearization:} Carefully selecting structured relinearization keys \( \mathbf{R}_1, \mathbf{R}_2 \) can minimize the additional noise introduced, reducing its impact on long-term computations.
\end{itemize}

\subsection{Security of VLWE-based Homomorphic Encryption}  
The security of the proposed homomorphic encryption scheme follows directly from the hardness of the Variety-LWE problem. Since the scheme does not introduce additional assumptions beyond the Variety-LWE framework, any attack on the encryption scheme would directly imply an attack on Variety-LWE, which has been proven to be at least as hard as solving multiple instances of Ideal-SVP.

\bibliographystyle{splncs04}
\bibliography{mybibliography}

\section*{Appendix: Comparison between VLWE and RLWE}

\paragraph{Computational Complexity: RLWE Requires Additional Encoding Overhead.} 
Recall that VLWE computations operate in a multivariate polynomial ring, where multiplication is performed coordinate-wise. The complexity of a single VLWE multiplication is $O(n d)$, assuming each variable is constrained by a degree-$d$ polynomial. In contrast, RLWE operates in a single-variable polynomial ring, requiring an encoding strategy to represent multiple variables. If RLWE were to simulate VLWE using multiple ciphertexts, the complexity increases to $O(n d \log d)$ due to FFT-based polynomial multiplication. Alternatively, using CRT-based encoding methods incurs additional encoding and decoding costs, making RLWE inherently less efficient than VLWE for high-dimensional computations.

\paragraph{Error Growth: VLWE Has More Controllable Error Propagation.} 
Since VLWE maintains independent polynomial constraints for each variable, error propagation is strictly linear in the number of variables, leading to an error growth of $O(n \sigma)$. In contrast, RLWE's error growth follows a quadratic trend, $O(d \sigma^2)$, due to polynomial multiplications introducing cross-terms. This means that, under repeated homomorphic operations, RLWE experiences a significantly faster noise expansion compared to VLWE, necessitating more frequent modulus switching and relinearization.

\paragraph{Security Assumptions: VLWE Reduces to Multiple Ideal-SVP Instances.} 
VLWE security is based on reductions to multiple independent Ideal-SVP instances, meaning that breaking VLWE requires solving several hard lattice problems simultaneously. RLWE, on the other hand, reduces to a single Ideal-SVP instance, making it theoretically more vulnerable in settings where Ideal-SVP instances can be correlated or attacked jointly. If RLWE were to encode multiple independent coordinates within a single ciphertext, the encoding structure might introduce hidden dependencies that could be exploited by algebraic or hybrid attacks.

\paragraph{Homomorphic Computation: VLWE is Better Suited for Vectorized Computations.} 
VLWE naturally supports vectorized homomorphic operations by design, where coordinate-wise multiplication aligns with operations frequently required in privacy-preserving machine learning and encrypted search. In contrast, RLWE requires additional transformations to achieve similar functionality, such as CRT-based batching, which imposes restrictions on the number of supported elements and introduces additional encoding overhead. As a result, VLWE provides a more direct and efficient framework for structured homomorphic encryption applications.

\end{document}